\documentclass[preprint,prd,amsmath,amssymb,floatfix,nofootinbib]{revtex4}
\usepackage[colorlinks=true,urlcolor=black]{hyperref}

\hypersetup{backref = true,pagebackref = true,hyperindex = true, colorlinks = true,
  breaklinks = true, urlcolor = blue, linkcolor = blue, bookmarks = true,
  bookmarksopen = true, citecolor=red}

\usepackage{bm}
\usepackage{color}
\usepackage{dcolumn}
\usepackage{graphics}
\usepackage{graphicx}
\usepackage{subfigure}
\usepackage{slashed}
\usepackage{placeins}

\usepackage{pdfpages}
\usepackage{eso-pic}
\usepackage{everyshi}
\usepackage{pdflscape}

\newcommand{\bea}{\begin{eqnarray}}
\newcommand{\eea}{\end{eqnarray}}
\newcommand{\be}{\begin{equation}}
\newcommand{\ee}{\end{equation}}

\newcommand{\ar}{a_s}

\begin{document}

\title{On Bjorken sum rule with analytic coupling
}
\author{I.R. Gabdrakhmanov$^{1}$, N.A Gramotkov$^{1,2}$, A.V.~Kotikov$^{1}$,  O.V.~Teryaev$^{1}$, D.A. Volkova$^{1,3}$  and I.A.~Zemlyakov$^{4}$}
\affiliation{
  $^1$Bogoliubov Laboratory of Theoretical Physics,
  Joint Institute for Nuclear Research, 141980 Dubna, Russia;\\
$^2$Moscow State University, 119991, Moscow, Russia\\
  $^3$Dubna State University,
  141980 Dubna, Moscow Region, Russia;\\
$^4$Department of Physics, Universidad Tecnica Federico Santa Maria, Avenida Espana 1680, Valparaiso, Chile}


\begin{abstract}

We present the results of \cite{Gabdrakhmanov:2024bje}, where good agreement was obtained
between  calculations within the framework of analytic QCD and experimental data on polarized
Bjorken sum rule. The photoproduction limit was also considered and a new representation of
the perturbative contribution to the polarized Bjorken sum rule was obtained.

\end{abstract}

\maketitle

\section{Introduction}

The polarized Bjorken sum rule (BSR) $\Gamma^{p-n}_1(Q^2)$ \cite{Bjorken:1966jh}, which is the difference between the first moments
of the spin structure functions (SF) of a proton and a neutron, is an important observable of Quantum Chromodynamics (QCD) \cite{Deur:2018roz}.
Its experimental data obtained in polarized deep inelastic scattering (DIS) are available on the scale of spacelike squared momenta $Q^2=-q^2$:
0.021 GeV$^2\leq Q^2 <$5 GeV$^2$ (see  \cite{Gabdrakhmanov:2024bje} and references therein).
In particular, recent results \cite{Deur:2021klh}  with small uncertainties  make BSR an observational activation for tests of various
generalizations of perturbative QCD (pQCD) in small $Q^2$ regions: $Q^2\leq 1$GeV$^2 $.

The standard approach to describing such quantities is pQCD in the $\overline{MS}$ renormalization scheme.
However, the presence of the Landau pole in the coupling constant ({\it couplant}) $\alpha_{\rm s}(Q^2)$ at $Q^2=\Lambda^2$ makes perturbation theory (PT)
inapplicable for $Q^2<1$ GeV. To correctly describe QCD observations in the nonperturbative region, an approach was developed that
eliminates the singularity at $Q^2=\Lambda^2$, called analytic PT (APT) \cite {ShS}.
The analytic couplant $A_{\rm MA}(Q^2)$ in the spacelike region can be represented as a spectral integral depending on the
PT order through the perturbative spectral function $r^{(i)}_{\rm pt}$:
\begin{equation}
	A^{(i)}_{\rm MA}(Q^2) 
	= \frac{i}{\pi} \int_{0}^{+\infty} \, 
	\frac{ d \sigma }{( \sigma + Q^2)} \, r^{(i)}_{\rm pt}( \sigma), \, \;\;
	r^{(i)}_{\rm pt}( \sigma)= {\rm Im} \; a_s^{(i)}(-\sigma - i \epsilon) \,.
	\label{disp_MA_LO}
\end{equation}
This approach is also called {\it Minimal Analytic} (MA) PT \cite{Cvetic:2008bn}.

Later, this approach was extended to arbitrary fractional degrees of couplant \cite{BMS1} and the so-called fractional analytic PT (FAPT) was formulated.
This extension involves a rather complex procedure, which is why the main results until recently were obtained mainly in the one-loop approximation.

Following \cite{Cvetic:2006mk}, we introduce derivatives (in $k$ order of PT) of the couplant
\be
\tilde{a}^{(k)}_{n+1}(Q^2)=\frac{(-1)^n}{n!} \, \frac{d^n a^{(k)}_s(Q^2)}{(dL)^n},~~a^{(k)}_s(Q^2)=\frac{\beta_0 \alpha^{(k)}_s(Q^2)}{4\pi}=\beta_0\,\overline{a}^{(k)}_s(Q^2).
\label{tan+1}
\ee
Here and below, $\beta_0$ is the zero coefficient of the QCD $\beta$-function:
\be
\beta(\overline{a}^{(k)}_s)=-{\left(\overline{a}^{(k)}_s\right)}^2 \bigl(\beta_0 + \sum_{i=1}^k \beta_i {\left(\overline{a}^{(k)}_s\right)}^i\bigr),
\label{bQCD}
\ee
where $\beta_i$ are known up to $i=4$ \cite{Baikov:2008jh}.

As was shown in \cite{Kotikov:2022swl}, the series of derivatives $\tilde{a}_{n}(Q^2)$ can serve as a replacement for the series of $\ar$ powers.
Indeed, although each differentiation reduces the $\ar$ degree, at the same time it creates an additional $\beta$-function containing the factor $\ar^2$.
In the leading approximation, the expressions $\tilde{a}_{n}(Q^2)$ and $\ar^{n}$ coincide, and in higher orders there is a one-to-one correspondence
between $\tilde{a}_{n}(Q ^2)$ and $\ar^{n}$ were established in \cite{Cvetic:2006mk,Cvetic:2010di} and extended to fractional powers in \cite{GCAK}.

In this article, we apply the inverse logarithm expansion for $A^{(i)}_{\rm MA}$, recently obtained \cite{Kotikov:2022sos} for an arbitrary PT order.
This approach is very convenient, because in one-loop order $A^{(1)}_{\rm MA}$ has a simple representation (see \cite{BMS1}),
and in higher orders $A^{(i)}_{\rm MA}$ are very close to $A^{(1)}_{\rm MA}$, especially for $Q^2 \to \infty$ and $Q^2 \to 0$,
where the difference between $A^{(i)}_{\rm MA}$ and $A^{(1)}_{\rm MA}$ disappears.
Moreover, for $Q^2 \to \infty$ and $Q^2 \to 0$ the corresponding derivatives $\tilde{A}^{(i)}_{\rm MA,n}$ for $n\geq 2$ tend to zero, and, therefore,
only the first term of the perturbative expansion makes a noticeable contribution.

\section{Bjorken sum rule}

The polarized BSR is defined as the difference between the polarized SFs of the proton and neutron, integrated over the entire interval $x$
\be
\Gamma_1^{p-n}(Q^2)=\int_0^1 \, dx\, \bigl[g_1^{p}(x,Q^2)-g_1^{n}(x,Q^2)\bigr].
\label{Gpn} 
\ee
and can be represented in the following form:
\be
\Gamma_1^{p-n}(Q^2)=
\frac{g_A}{6} \, \bigl(1-D_{\rm BS}(Q^2)\bigr) +\frac{\hat{\mu}_4 M^2}{Q^{2}+M^2} \, ,
\label{Gpn.mOPE} 
\ee
where $g_A$=1.2762 $\pm$ 0.0005 \cite{PDG20} is the axial charge of the nucleon, $(1-D_{BS}(Q^2))$ is the contribution of the leading twist (twist-2),
and $(\hat{\mu}_4 M^2)/(Q^{2}+M^2)$ is so-called "massive" representation for twist-4 (see \cite{Teryaev:2013qba}).

The perturbative expansion up to the $k$-th order has the form
\be
D^{(1)}_{\rm BS}(Q^2)=\frac{4}{\beta_0} \, \tilde{a}^{(1)}_1,~~D^{(k\geq2)}_{\rm BS}(Q^2)=
\frac{4}{\beta_0} \, \left(\tilde{a}^{(k)}_{1}+\sum_{m=2}^k\tilde{d}_{m-1}\tilde{a}^{(k)}_{m}
\right),
\label{DBS.1} 
\ee
where $\tilde{d}_1$, $\tilde{d}_2$ and $\tilde{d}_3$ are known from direct calculations (see, for example, \cite{Chen:2006tw}).
The coefficient $\tilde{d}_4$ is not known, but there is an estimate for it, found in \cite{Ayala:2022mgz}.

Here and below we will take the number of active quark flavors $f=3$. Thus,
\footnote{
  The coefficients $\beta_i$ $(i\geq 0)$ of QCD $\beta$ function (\ref{bQCD}) and, as a consequence, the couplant $\alpha_s(Q^2)$
  depend on the number $f$ of active flavors, and each new quark is included in the game at the threshold level $Q^2_f$
  in accordance with \cite{Chetyrkin:2005ia}.
The corresponding QCD parameters $\Lambda^{(f)}$ in N$^i$LO of PT were obtained in \cite{Chen:2021tjz}.
}
\bea
\tilde{d}_1=1.59,~~\tilde{d}_2=2.73,
~~\tilde{d}_3=8.61,~~\tilde{d}_4=21.52 \, .
\label{td123} 
\eea

Moving from the series $D_{\rm BS}(Q^2)$ to the same series, but along $\tilde{A}^{(k)}_{\rm MA,n}$:
\be
D^{(1)}_{\rm MA,BS}(Q^2)=\frac{4}{\beta_0} \, A_{\rm MA}^{(1)},
D^{k\geq2}_{\rm{MA,BS}}(Q^2) =\frac{4}{\beta_0} \, \Bigl(A^{(1)}_{\rm MA}
+ \sum_{m=2}^{k} \, \tilde{d}_{m-1} \, \tilde{A}^{(k)}_{\rm MA,\nu=m} \Bigr)\,,
\label{DBS.ma} 
\ee
we obtain a replacement (\ref{Gpn.mOPE}) within the framework of APT in the form
\be
\Gamma_{\rm{MA},1}^{p-n}(Q^2)=
\frac{g_A}{6} \, \bigl(1-D_{\rm{MA,BS}}(Q^2)\bigr) +\frac{\hat{\mu}_{\rm{MA},4}M^2}{Q^{2}+M^2}.
\label{Gpn.MA} 
\ee

\section{Results}

\begin{table}[h!]
\begin{center}
\begin{tabular}{|c|c|c|c|}
\hline
& $M^2$ for (\ref{Gpn.MA})& $\hat{\mu}_{\rm{MA},4}$& $\chi^2/({\rm d.o.f.})$  for \\
& (for (\ref{Gpn.MAn})) &for (\ref{Gpn.MA}) &  (\ref{Gpn.MA}) (for (\ref{Gpn.MAn})) \\
 \hline
 LO & 1.631 $\pm$ 0.301 (0.576 $\pm$ 0.046) & -0.166 $\pm$ 0.001 & 0.572 (0.575) \\
 \hline
 NLO & 1.545 $\pm$ 0.287  (0.464 $\pm$ 0.039) & -0.155 $\pm$ 0.001 & 0.586 (0.590) \\
 \hline
 N$^2$LO & 1.417 $\pm$ 0.261 (0.459 $\pm$ 0.038) &-0.156 $\pm$ 0.002 & 0.617 (0.584) \\
 \hline
 N$^3$LO & 1.429 $\pm$ 0.248  (0.464 $\pm$ 0.039) & -0.157 $\pm$ 0.002 & 0.629 (0.582) \\
   \hline
N$^4$LO & 1.462 $\pm$ 0.253  (0.465 $\pm$ 0.039) &-0.157 $\pm$ 0.001 & 0.625  (0.584) \\
 \hline
\end{tabular}
\end{center}
\caption{%
Fitting parameters.
}
\label{Tab:BSR1}
\end{table}

\begin{figure}[t]
\centering
\includegraphics[width=0.98\textwidth]{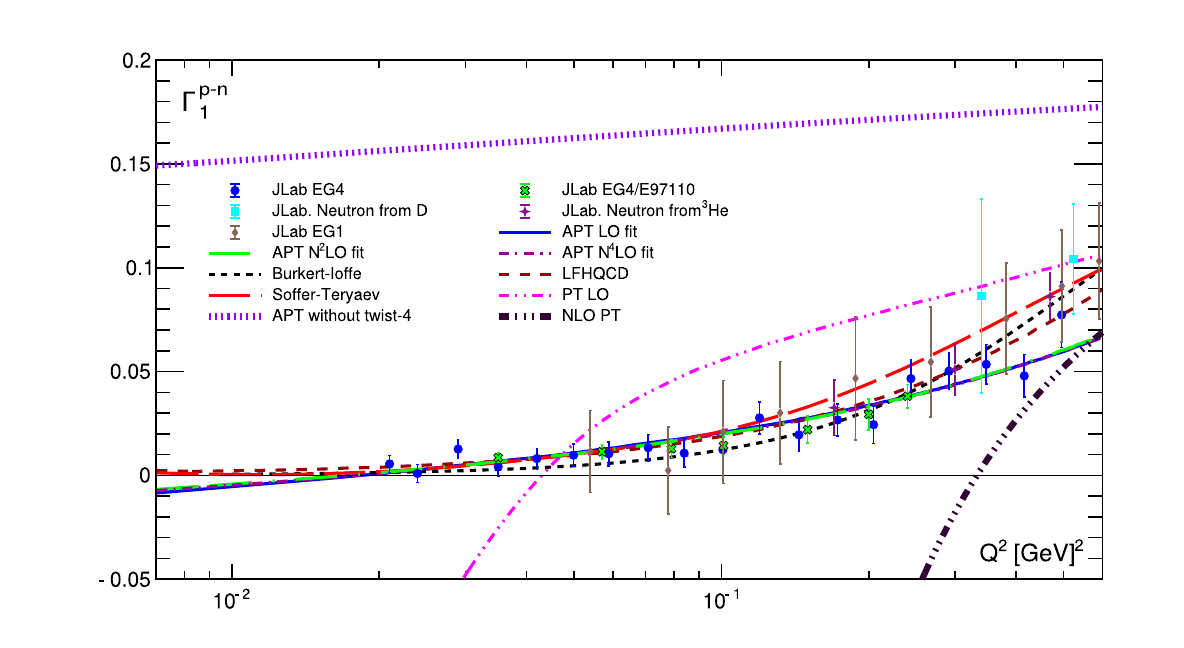}
\caption{
  \label{fig:low}
  Results for $\Gamma_1^{p-n}(Q^2)$ in the first five orders of APT, obtained from experimental data fits at $Q^2 <$0.6 GeV$^2$
}
\end{figure}

The analysis of experimental data was carried out taking into account only statistical errors, the results are
presented in Tab.\ref{Tab:BSR1} and shown in Fig.\ref{fig:low}.
The quantities were fitted as $Q^2$-independent using twist-2 from (\ref{DBS.1}) and (\ref{DBS.ma}) for regular PT and APT, respectively.
In the PT case, the results obtained are not able to describe the experimental data, and the discrepancy increases with the PT order,
which was discussed in detail in \cite{Pasechnik:2008th,Ayala:2017uzx,Gabdrakhmanov:2023rjt,Gabdrakhmanov:2024bje}.
Here we will limit ourselves only to the analyses for small $Q^2$. A more general case can be found in our paper \cite{Gabdrakhmanov:2024bje}.

As can be seen in Fig.\ref{fig:low}, our results in different APT orders are barely distinguishable.
The high quality of fits is confirmed by the small values of the corresponding $\chi^2/({\rm d.o.f.})$ (see Tab.\ref{Tab:BSR1}).
At $Q^2\leq 0.3$GeV$^2$, good agreement with phenomenological models was found \cite{Burkert:1992tg}.
However, as $Q^2$ increases, our results are lower than the predictions of phenomenological models,
and for $Q^2\geq 0.5$GeV$^2$ are lower than experimental points.

As can be seen, the results for $\Gamma_{\rm{MA},1}^{p-n}(Q^2)$ take unphysical negative values for $Q^2 <$0.02 GeV$^2$.
The reason for this phenomenon can be clarified by considering photoproduction.\\

{\bf Photoproduction.}
To understand the $\Gamma_{\rm{MA},1}^{p-n}(Q^2\to 0)<0$ problem, let us turn to the case of photoproduction, where
  \be
  A^{(k)}_{\rm MA}(Q^2=0)
  =1,~~ \tilde{A}^{(k)}_{{\rm MA},m}=0,~~\mbox{for}~~m>1
\label{Akm} 
\ee
and, thus,
  \be
D_{\rm MA,BS}(Q^2=0)=\frac{4}{\beta_0} 
~~\mbox{and therefore,}~~
\Gamma_{\rm{MA},1}^{p-n}(Q^2=0)=
\frac{g_A}{6} \, \bigl(1-\frac{4}{\beta_0}\bigr) +\hat{\mu}_{\rm{MA},4}
\,.
\label{Gpn.MA.Q0} 
\ee

The finiteness of the cross section in the limit of real photons leads to the requirement (see \cite{Teryaev:2013qba}):
\be
\Gamma_{\rm{MA},1}^{p-n}(Q^2=0)=0
~~\mbox{and therefore,}~~
\hat{\mu}^{php}_{\rm{MA},4}=-\frac{g_A}{6} \, \bigl(1-\frac{4}{\beta_0}\bigr).
\label{mu.GDH} 
\ee
For $f=3$, we have
\be 
\hat{\mu}^{php}_{\rm{MA},4}=-0.118
~~\mbox{and therefore,}~~
|\hat{\mu}^{php}_{\rm{MA},4}|< |\hat{\mu}_{\rm{MA},4}|,
\label{mu.GDH} 
\ee
as can be seen in Tab.\ref{Tab:BSR1}.

Thus, the results of our analysis are not consistent with the finiteness of the cross section in the limit of real photons.
\footnote{
  Note that the results for $\hat{\mu}_{\rm{MA},4}$ were obtained taking into account only statistical errors. When adding
  systematic errors, the results for $\hat{\mu}^{php}_{\rm{MA},4}$ and $\hat{\mu}_{\rm{MA},4}$ are absolutely consistent with
  each other, but the predictive power of such an analysis is small.}
This violation leads to negative values of $\Gamma_{\rm{MA},1}^{p-n}(Q^2\to 0)$.\\

{\bf Gerasimov-Drell-Hearn and Burkhardt-Cottingham sum rules.}
To improve the analysis of experimental data at small $Q^2$, we add the contribution of twist-6 in
the ``massive'' representation, similar to twist-4 in (\ref{Gpn.mOPE}).
Moreover, let us add into consideration the Gerasimov-Drell-Hearn and Burkhadra-Cotingham sum rules,
which leads to the following results (see \cite{Teryaev:2013qba,Soffer:1992ck}):
  \be
  \frac{d}{dQ^2} \Gamma_{\rm{MA},1}^{p-n}(Q^2=0)= G,~~G=\frac{\mu^2_n-(\mu_p-1)^2}{8M_p^2}=0.0631\,,
  \label{GDH} 
  \ee
  where $\mu_n=-1.91$ and $\mu_p=2.79$ are the magnetic moments of the neutron and proton, respectively,
  and $M_p$ = 0.938 GeV is the mass of the nucleon. Note that $G$ is small.

  From (\ref{tan+1}) we have that
  \be
Q^2\frac{d}{dQ^2} \tilde A_n(Q^2) \sim \tilde A_{n+1}(Q^2)\,.
  \label{An} 
  \ee
  and thus,
  \be
Q^2\frac{d}{dQ^2}  \tilde A_n(Q^2) \to 0\,,~~\mbox{but}~~
\frac{d}{dQ^2}  \tilde A_n(Q^2\to 0) \to \infty \,.
  \label{AnQ0} 
  \ee

  Therefore, after applying the derivative $d/dQ^2$ to $\Gamma_{\rm{MA},1}^{p-n}(Q^2)$ each term in $D_{\rm MA,BS}(Q^ 2)$
  becomes divergent as $Q^2 \to 0$.
  To achieve finiteness at $Q^2 \to 0$ of the left-hand side (\ref{GDH}), let us assume a connection between the contributions
  of twist-2 and twist-4, which will lead to the appearance of a new term
\be
-\frac{g_A}{6}\,D_{\rm MA,BS}(Q^2) +  \frac{\hat{\mu}_{\rm{MA},4}M^2}{Q^{2}+M^2}\,D_{\rm MA,BS}(Q^2)\,,
  \label{DD} 
  \ee
  which can be made finite for $Q^2 \to 0$.
  The form (\ref{DD}) is equivalent to the following modification of the expression $\Gamma_{\rm{MA},1}^{p-n}(Q^2)$ in (\ref{Gpn.MA}):
  \be
\Gamma_{\rm{MA},1}^{p-n}(Q^2)=
\, \frac{g_{A}}{6} \, \bigl(1-D_{\rm{MA,BS}}(Q^2) \cdot \frac{Q^2}{Q^2+M^2}\bigr) +
\frac{\hat{\mu}_{\rm{MA},4}M^2}{Q^{2}+M^2}+\frac{\hat{\mu}_{\rm{MA},6}M^4}{(Q^{2}+M^2)^2}.~~
\label{Gpn.MAn} 
\ee

The finiteness of the cross section for real photons now leads to \cite{Teryaev:2013qba}
\be
\Gamma_{\rm{MA},1}^{p-n}(Q^2=0)=0=\frac{g_{A}}{6}+\hat{\mu}_{\rm{MA},4}+\hat{\mu}_{\rm{MA},6}\,
\label{Gpn.MAnQ0} 
\ee
and thus we have
\be
\hat{\mu}_{\rm{MA},4}+\hat{\mu}_{\rm{MA},6}=-\frac{g_{A}}{6} \approx - 0.21205\,.
\label{Gpn.MAnQ0.1} 
\ee

From Eqs. (\ref{Gpn.MA.Q0}) and (\ref{Gpn.MAn}) and conditions (\ref{GDH}) we obtain that

\be
-\frac{2g_{A}}{3\beta_0} - \hat{\mu}_{\rm{MA},4}-2\hat{\mu}_{\rm{MA},6}=G\,M^2\,.
\label{Gpn.MAnQ0.2} 
\ee

Substituting $f=3$ (i.e. $\beta_0=9$), we have
\be
\hat{\mu}_{\rm{MA},4}+2\hat{\mu}_{\rm{MA},6}=
-G\,M^2-\frac{2g_{A}}{27} \approx -G\,M^2- 0.0945\,.
\label{Gpn.MAnQ0.3} 
\ee

Combining (\ref{Gpn.MAnQ0}) and (\ref{Gpn.MAnQ0.3}) together we arrive at the following results:
\bea
\hat{\mu}_{\rm{MA},6} =-G\,M^2+\frac{5g_{A}}{54}= -G\,M^2+0.1182,\nonumber \\
\hat{\mu}_{\rm{MA},4} = -\frac{g_{A}}{6} -\hat{\mu}_{\rm{MA},6}= G\,M^2-\frac{7g_{A/V}}{27}=  G\,M^2-0.3309\,.
\label{Gpn.MAnQ0.4} 
\eea
Because $G$ is small, then $\hat{\mu}_{\rm{MA},4}<0$ and $\hat{\mu}_{\rm{MA},6}
>0$.

\begin{figure}[t]
\centering
\includegraphics[width=0.98\textwidth]{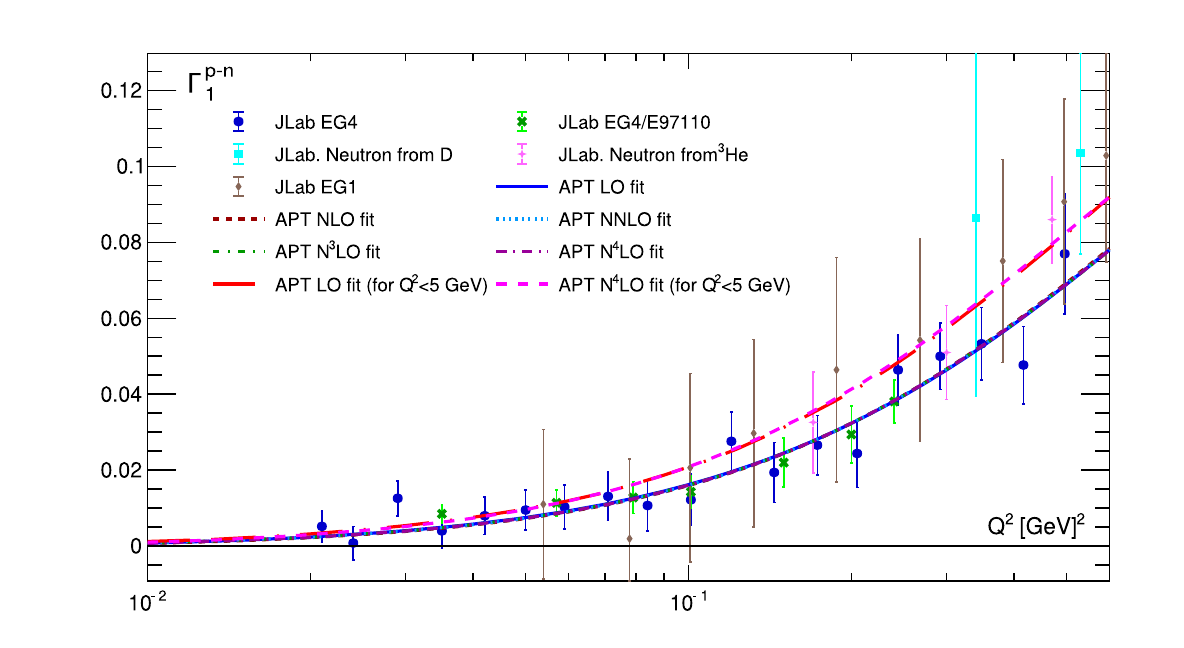}
\caption{
  \label{fig:low1}
Same as in Fig.\ref{fig:low}, but for the modification (\ref{Gpn.MAn})
}
\end{figure}

Results of fitting theoretical predictions based on the expression (\ref{Gpn.MAn}) with $\hat{\mu}_{\rm{MA},4}$
and $\hat{\mu}_{\rm{MA },6}$ obtained (\ref{Gpn.MAnQ0.4}), are presented in Tab.\ref{Tab:BSR1} and Fig.\ref{fig:low1}.

\section{Conclusions}

We gave a brief overview of the results of the paper \cite{Gabdrakhmanov:2024bje}, where we considered the
Bjorken sum rule $\Gamma_{1}^{p-n}(Q^2)$ within the framework of APT and perturbative QCD in the first five
orders of PT and obtained the results, similar to those obtained in previous studies
\cite{Pasechnik:2008th,Ayala:2017uzx,Gabdrakhmanov:2023rjt}.
Perturbative QCD significantly diverges from experimental data.
For some $Q^2$ its results become negative, because higher order corrections are included in the twist-2
part with a negative sign.
The use of APT leads to good agreement with experimental data when using the ``massive'' variant (\ref{Gpn.MA})
for twist-4 contributions.

However, when examining the $Q^2\to 0$ limit, we discovered a discrepancy between the data and APT.
The results of fits performed in the region of small $Q^2$ lead to $\Gamma_{\rm{MA},1}^{p-n}(Q^2\to 0) <0$,
which contradicts the finiteness of the cross section in the limit of real photons,
requiring $\Gamma_{\rm{MA},1}^{p-n}(Q^2\to 0) =0$.

To solve this problem, in \cite{Gabdrakhmanov:2024bje} we introduced the infrared modification (\ref{Gpn.MAn})
of the operator expansion $\Gamma_{\rm{MA},1}^{p-n}(Q^2)$ and came to good agreement with a complete set of
experimental data  for the Bjorken sum rule $\Gamma_{\rm{MA},1}^{p-n}(Q^2)$ up to the photoproduction limit
$Q^2 \to 0$. Good agreement with phenomenological models was also found.

{\bf Acknowledgments}~~Authors are grateful to Alexandre P. Deur for
information about new experimental data in Ref. \cite{Deur:2021klh} and discussions.
Authors thank Andrei Kataev and
Nikolai Nikolaev for careful discussions.
This work was supported in part by the Foundation for the Advancement of Theoretical
Physics and Mathematics “BASIS”.
One of us (I.A.Z.)
is supported by the Directorate of
Postgraduate Studies of the Technical University of Federico Santa Maria.




\end{document}